\documentclass[aps,prl,twocolumn, superscriptaddress,showpacs]{revtex4-1}

\usepackage[T1]{fontenc}
\usepackage[latin1]{inputenc}
\usepackage[dvips]{graphicx,color}
\usepackage{rotating}
\usepackage{bm}        
\usepackage{amssymb}   
\usepackage{amsmath}
\def\jcp#1#2#3{J.~Chem.~Phys.~{\bf #1},\ #2\ (#3)}

\def\pra#1#2#3{Phys.~Rev.~A~{\bf #1},\ #2\ (#3)}
\def\prl#1#2#3{Phys.~Rev.~Lett.~{\bf #1},\ #2\ (#3)}

\def\k1{k_1}
\def\k2{k_2}
\def\q1{q_1}
\def\q2{q_2}

\def\({\left (}
\def\){\right )}
\def\[{\left [}
\def\]{\right ]}

\newcommand{\beq}{\begin{equation}}
\newcommand{\eeq}{\end{equation}}

\begin{document}
\date{\today}
\title{Ultracold spin-polarized mixtures of $^2\Sigma$ molecules with $S$-state atoms: \\ Collisional stability and implications for sympathetic cooling}

\author{T. V. Tscherbul}
\affiliation{Harvard-MIT Center for Ultracold Atoms, Cambridge, Massachusetts 02138}
\affiliation{ITAMP, Harvard-Smithsonian Center for Astrophysics, Cambridge, Massachusetts 02138}
\author{J. K{\l}os}
\affiliation{Department of Chemistry and Biochemistry, University of Maryland, College Park, Maryland 20742}
\author{A. A. Buchachenko}
\affiliation{Department of Chemistry, Moscow State University, Moscow 119991, Russia}

\begin{abstract}
The prospects of sympathetic cooling of polar molecules with magnetically co-trapped alkali-metal atoms are generally considered poor due to strongly anisotropic atom-molecule interactions leading to  large spin relaxation rates. Using rigorous quantum scattering calculations based on {\it ab initio} interaction potentials, we show that inelastic spin relaxation in low-temperature collisions of CaH($^2\Sigma$) molecules with Li and Mg atoms occurs at a slow rate despite the strongly anisotropic interactions. This unexpected result, which we rationalize using multichannel quantum defect theory, opens up the possibility of sympathetic cooling of polar $^2\Sigma$ molecules with alkali-metal and alkaline-earth atoms in a magnetic trap.
\end{abstract}

\maketitle

\clearpage
\newpage


The experimental realization of molecular ensembles cooled to temperatures below 1 K has opened up a multitude of fascinating research directions in physics and chemistry \cite{NJP}. The unique properties of ultracold molecular gases such as their long-range, anisotropic interactions may be used to implement quantum logic gates \cite{Andre}, create and explore novel phases of quantum matter \cite{Barbara}, and study non-equilibrium dynamics and many-body localization phenomena \cite{Roman10}.
Recent experimental and theoretical work has demonstrated the possibility of controlling chemical reactions in an ultracold gas of KRb molecules by applying external electric fields and confining the reactants in low dimensions \cite{KRb1,KRb2}, opening up a vast new area of research in physical chemistry \cite{NJP}.

A variety of experimental techniques has been developed for cooling and trapping polar molecules \cite{NJP}. Indirect cooling techniques such as photo- and magnetoassociation \cite{NJP}
assemble ultracold molecules from precursor atoms using laser light and oscillating magnetic fields.
While capable of producing high phase-space density gases of ground-state polar molecules, the indirect techniques are currently limited in scope to $\Sigma$-state alkali-metal dimers \cite{NJP}. Alternatively, molecules can be cooled directly via thermal contact with cryogenic He buffer gas (buffer-gas cooling \cite{NJP,Weinstein:98}) or molecular beam deceleration \cite{StarkDeceleration}. The direct cooling techniques are ideally suited for the production of diverse classes of polar molecules required for specific applications in quantum simulation \cite{NJP} and precision measurements \cite{OH,YbF}.
However, these techniques tend to produce cold ($T>10$ mK) rather than ultracold molecules, necessitating an additional stage of cooling to reach ultralow temperatures.

 Sympathetic cooling is arguably the most straightforward way of cooling molecules below 10 mK by bringing them in thermal contact with a reservoir of ultracold atoms such as the alkali-metal atoms \cite{RbOH,RbNH,RbNH3}. As this technique does not rely on specific details of molecular structure, it can potentially lead to the production of a wide array of ultracold molecules.  Molecular cooling and trapping experiments typically employ static electromagnetic traps, which confine molecules in their low-field-seeking Zeeman states \cite{Weinstein:98}. Inelastic collisions with buffer-gas atoms lead to spin relaxation and trap loss \cite{Roman03}, and a large ratio of elastic to inelastic collision rates ($\gamma>100$) is required for efficient cooling \cite{NJP}.  Previous theoretical work has shown that spin relaxation in collisions of alkali-metal atoms with OH \cite{RbOH}, NH \cite{RbNH}, and NH$_3$ \cite{RbNH3} is extremely fast ($\gamma\sim1$). Based on these results, the alkali-metal atoms were judged unsuitable as collision partners for sympathetic cooling of molecules, and
attention turned to the alkaline-earth \cite{MgNH}, and rare-gas \cite{Barker} atoms, which lack magnetic moments and are thus challenging to accumulate in a magnetic trap. Spin-polarized nitrogen atoms have favorable collisional properties with NH molecules \cite{N-NH} but their use in sympathetic cooling experiments is limited by a lack of efficient detection techniques.



In this Letter,  we explore the possibility of sympathetic cooling of diatomic molecules with one unpaired electron by collisions with spin-polarized alkali-metal and alkali-earth atoms in a magnetic trap. Using converged quantum scattering calculations based on {\it ab initio} interaction potentials, we demonstrate that spin relaxation in collisions of CaH($^2\Sigma$)  molecules with Li and Mg atoms occurs slowly {\it despite} the fact that Li-CaH and Mg-CaH interactions are extremely strong and anisotropic. To explain this surprising result, we map out the dependence of the inelastic cross section on the potential anisotropy using multichannel quantum defect theory. Our analysis shows that spin relaxation in low-temperature collisions of $^2\Sigma$ molecules can be slow even in systems with strongly anisotropic interactions such as Li-CaH, suggesting that sympathetic cooling of $^2\Sigma$ molecules with alkali-metal atoms in a magnetic trap is likely to be successful.






We begin by outlining our theoretical approach. The Hamiltonian of the collision complex formed by a $^2\Sigma$ molecule A and an $S$-state atom B is ($\hbar=1$) \cite{jcp10,Roman04}
\begin{multline}\label{H}
\hat{H} = -\frac{1}{2\mu R}\frac{\partial^2}{\partial R^2}R + \frac{\hat{l}^2}{2\mu R^2} + \sum_{S,\Sigma} V_{S\Sigma}(R,r,\theta) |S\Sigma\rangle \langle S\Sigma| \\ + \hat{H}_\text{A} + \hat{H}_\text{B}  - \sqrt{\frac{24\pi}{5}} \frac{\alpha^2}{R^3}  \sum_{q}Y^\star_{2q}(\hat{R})[\hat{S}_\text{A}\otimes\hat{S}_\text{B}]^{(2)}_q
\end{multline}
where $\mu$ is the reduced mass, $R$ is the atom-molecule separation,
$r$ is the internuclear distance, $\theta$ is the angle between the Jacobi
vectors $\mathbf{R}$ and $\mathbf{r}$, $\hat{l}$ is the orbital
angular momentum for the collision, $V_{S\Sigma}(R,r,\theta)$ is the
potential energy surface (PES) for the atom-molecule interaction, and
$\hat{S}=\hat{S}_\text{A}+\hat{S}_\text{B}$ is the total spin. The Hamiltonian of the
$^2\Sigma$ molecule is given by $\hat{H}_\text{A} = B_e\hat{N}^2 + \gamma \hat{N}\cdot\hat{S}_\text{A} + 2\mu_0 B \hat{S}_{\text{A}z}$,
where $\hat{N}$ is the rotational angular momentum of the molecule, ${B}$ is the magnetic field strength,
$\hat{S}_{\text{A}z}$ is the projection of molecular spin $\hat{S}_\text{A}$ on the field axis, $B_e$ is the rotational constant, $\gamma$ is the spin-rotation constant \cite{Constants}, and $\mu_0$ is the Bohr magneton. The atomic Hamiltonian is given by $\hat{H}_\text{B} = 2\mu_0 B \hat{S}_{\text{B}z}$ and the term proportional to $R^{-3}$ represents the magnetic dipole interaction. Both of these terms are absent for the Mg atom bearing no magnetic moment.
In this work, we are interested in collisions of CaH molecules with Li atoms initially in their fully spin-polarized Zeeman states $M_{S_\text{A}}=M_{S_\text{B}}=1/2$, where $M_{S_\text{A}}$ and $M_{S_\text{B}}$ are the projections of $\hat{S}_\text{A}$ and $\hat{S}_\text{B}$ on the magnetic field axis. We can therefore neglect the weak magnetic dipole and spin-rotation couplings between basis states of different $S$ \cite{N-NH}, and include only the $S=1$ (triplet) Li-CaH PES in scattering calculations.

The interaction PESs for Li-CaH and Mg-CaH were evaluated {\it ab initio} \cite{MOLPRO} using a highly correlated open-shell coupled cluster CCSD(T) method and large quadruple-$\zeta$ correlation-consistent basis sets augmented by atomic-centered diffuse and $R$-centered bond functions. The CaH fragment was kept frozen at equilibrium ($r_e = 3.803a_0$). To study the dependence of the Li-CaH interaction energy on the CaH stretching coordinate $r$,
we calculated the PES over the range of $r= 3.0{-}5.7a_0$. We found that the interaction potential becomes slightly deeper with increasing $r$, but retains its repulsive wall. This demonstrates that the chemical reaction Li + CaH $\to$ LiH + Ca, which is exothermic by $\sim$0.6 eV, is forbidden when both the reactants are spin-polarized. We note that the reaction has been observed to occur rapidly at cryogenic temperatures with {\it unpolarized} reactants \cite{DAMOP}.

\begin{figure}[t]
        \centering
        \includegraphics[width=0.4\textwidth, trim = 0 0 0 0]{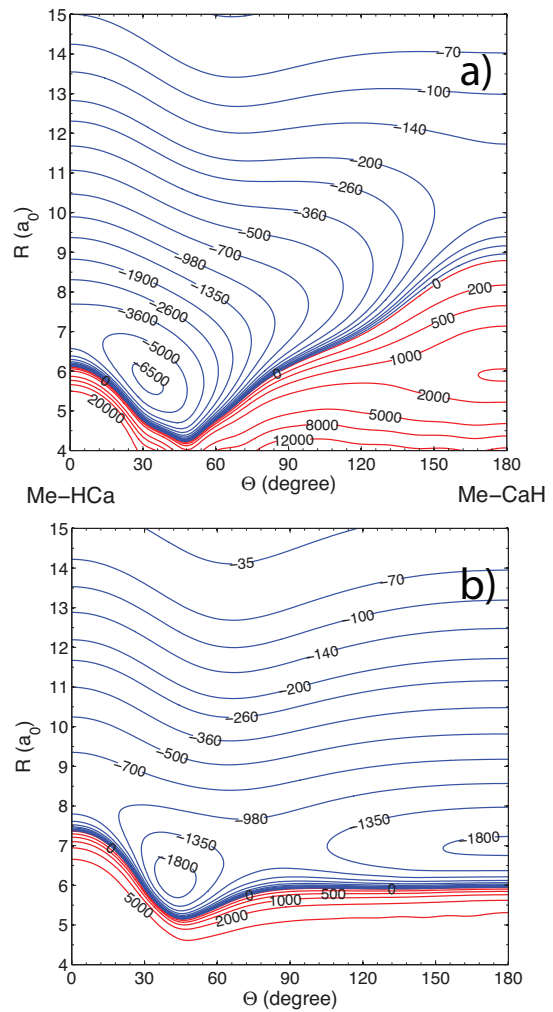}
        \renewcommand{\figurename}{Fig.}
        \caption{Contour plots of the {\it ab initio} PESs for Li-CaH ($S=1$) (a) and Mg-CaH (b). Energies are in units of cm$^{-1}$. Collinear arrangements of atoms for $\theta=0^\circ$ and $180^\circ$ are  indicated between panels, where Me = Li, Mg.  }\label{fig:pes}
\end{figure}







A contour plot of our {\it ab initio} PES for Li-CaH is shown in Fig. \ref{fig:pes}a. The global minimum is located at $R=5.6$ $a_0$, $\theta=35^\circ$ and is 7063 cm$^{-1}$ deep. Two saddle points occur in the linear Li-CaH (Li-HCa) configurations at $R=10.7a_0$ ($7.0a_0$) with the corresponding well depths of $-158.5$ and $-4384$ cm$^{-1}$, respectively.  This demonstrates that the Li-CaH interaction is not only strong, but also strikingly anisotropic.  For Mg-CaH, in addition to the global minimum of 2110 cm$^{-1}$ at $R=6.08 a_0$, $\theta=42^\circ$, we find a local minimum at the linear Mg-CaH arrangement at $R=7.0 a_0$ with the well depth of 1855 cm$^{-1}$. As shown in Fig. \ref{fig:pes}, both the Li-CaH and Mg-CaH PESs have similar topology at long range, but differ significantly at small $R$, with the Li-CaH interaction being significantly more anisotropic.

\begin{figure}[t]
        \centering
        \includegraphics[width=0.4\textwidth, trim = 0 0 0 0]{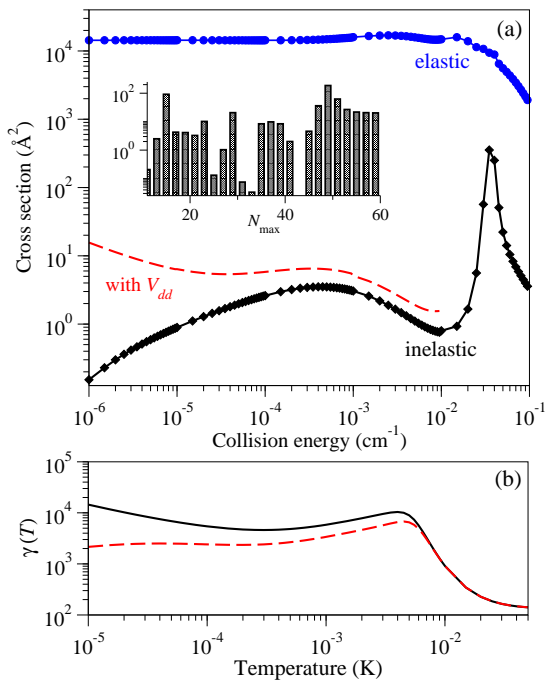}
        \renewcommand{\figurename}{Fig.}
        \caption{(a) Cross sections for spin relaxation in Li-CaH collisions calculated as functions of collision energy for $B=0.1$~T. Full lines and symbols -- calculations without the magnetic dipole interaction, dashed line -- calculations including the magnetic dipole interaction. (b) Thermally averaged ratio of elastic to inelastic collision rates for Li-CaH as a function of temperature. The inset shows the convergence of the inelastic cross section at $\epsilon=0.05$ cm$^{-1}$ with the maximum number of rotational channels included in the basis set ($N_\text{max}$).}\label{fig:LiCaH}
\end{figure}

In order to assess the prospects for sympathetic cooling of CaH molecules with Li and Mg atoms in a magnetic trap, we solve the quantum collision problem specified by the Hamiltonian (\ref{H}) numerically using a close-coupling (CC) approach in the body-fixed coordinate frame \cite{jcp10}.  Due to the large anisotropy of Li-CaH and Mg-CaH interactions, a large number of rotational channels must be included in the basis set to obtain converged results. As shown in the inset of Fig. \ref{fig:LiCaH}a, the cross sections oscillate dramatically with increasing the size of the rotational basis set until convergence is reached at $N_\text{max}=55$. Four total angular momentum states ($J = 0.5 - 4.5$) were included in the basis to produce the cross sections converged to $<$10~\%. The maximum number of channels was 3250 for the total angular momentum projection $M=1/2$. We note that the same problem reformulated in the space-fixed representation \cite{Roman04} with $N_\text{max}=55$, $l_\text{max}=55$ would amount to 234136 channels, making it computationally intractable. 





\begin{figure}[t]
        \centering
        \includegraphics[width=0.385\textwidth, trim = 0 0 0 0]{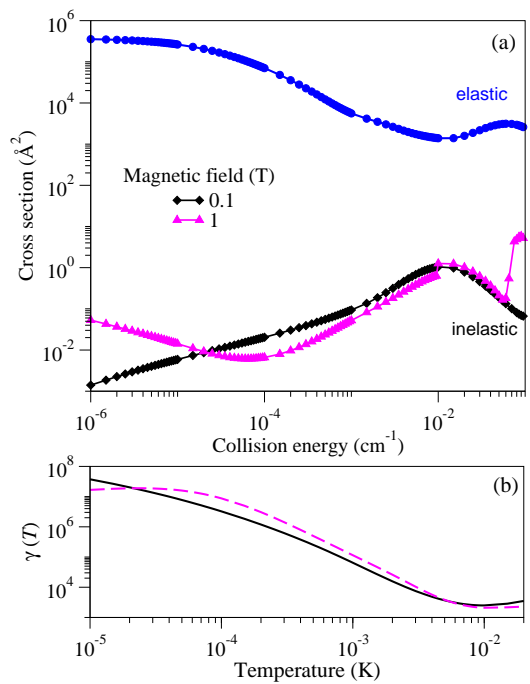}
        \renewcommand{\figurename}{Fig.}
        \caption{(a) Cross sections for spin relaxation in Mg-CaH collisions calculated as functions of collision energy for $B=0.1$~T (diamonds) and 1 T (triangles); (b) Thermally averaged ratio of elastic to inelastic collision rates for Mg-CaH vs temperature for $B=0.1$ T (full line) and 1 T (dashed line).}\label{fig:MgCaH}
\end{figure}

Figure \ref{fig:LiCaH}(a) shows the cross sections for spin relaxation and elastic scattering in Li-CaH collisions as functions of collision energy calculated for a fixed value of the magnetic field ($B = 0.1$ T). The inelastic cross sections remain small (below 0.5 {\AA}) over the entire range of collision energies considered except in the vicinity of a  shape resonance at  $\epsilon \sim 0.05$ cm$^{-1}$. 
The magnetic dipole interaction has a dramatic effect on inelastic cross sections at low collision energies; the cross sections assume their characteristic threshold behavior at a higher $\epsilon$ than they would in the absence of the interaction. In this regime, our calculated spin relaxation rates for Li-CaH are similar in magnitude to those observed for spin-polarized alkali-metal atoms \cite{Hulet}. At $\epsilon >0.001$ cm$^{-1}$, the magnetic dipole interaction has little effect on the inelastic cross sections, which are dominated instead by the intramolecular spin-rotation interaction. The Mg-CaH interaction is more weakly anisotropic and the cross sections for spin relaxation in Mg-CaH collisions shown in Fig. \ref{fig:MgCaH}a are smaller than for Li-CaH.  The ratios of elastic-to-inelastic collision rates for Mg-CaH displayed in Fig. \ref{fig:MgCaH}b are above $10^3$ at all temperatures, suggesting excellent prospects for sympathetic cooling of CaH molecules with Mg atoms.   

Given the extremely large anisotropy of the Li-CaH and Mg-CaH interactions, it is remarkable that the inelastic cross sections shown in Figs. \ref{fig:LiCaH} and \ref{fig:MgCaH} are so small in absolute magnitude. In order to gain insight into the mechanism of spin relaxation in strongly anisotropic collision systems, we employ the powerful formalism of multichannel quantum defect theory (MQDT) \cite{MQDT}. We begin by partitioning the full $K$-matrix into closed and open-channel blocks \cite{MQDT} using a minimal basis set consisting of two open ($|N=0,M_{S_\text{A}}=\pm 1/2\rangle$) and two closed ($|N=1,M_{S_\text{A}}=\pm1/2\rangle$) channels. In the absence of the magnetic dipole interaction, there is no direct coupling between the open channels \cite{Roman03}, and we find $K^\text{oo}=0$ and $K^\text{oc}=K^\text{co}=\text{diag}(K_1,K_1)$, where $K_1$ quantifies the strength of the anisotropic coupling between the $N=0$ and $N=1$ rotational states. 
 The closed-closed block of the $K$-matrix is given by $K^\text{cc}  = \left(\begin{array}{cc} K_{bb} & K_\gamma \\ K_\gamma & K_{bb} \end{array}\right)$, where $K_{bb}$ is the closed-channel level shift \cite{MQDT} and $K_\gamma$ is the spin-rotation coupling between the closed channels. We note that $K_\gamma$ is small, and can thus be approximated as $K_\gamma = -\gamma {\partial \nu_b}/{\partial \epsilon}$, where $\nu_b$ is the quantum defect \cite{MQDT}.

 Having defined our model $4\times 4$ $K$-matrix, we obtain the reduced $K$-matrix using $K^\text{red} = K^\text{oo} + K^\text{oc} [\tan \nu_b + K^\text{cc}]^{-1} K^\text{co}$ \cite{MQDT}:
 \begin{equation}\label{K}
K^\text{red} = \frac{-1}{D} \left(\begin{array}{cc} K_1^2(\tan\nu_b + K_{bb}) & -K_\gamma K_1^2  \\
                                 -K_\gamma K_1^2  & K_1^2(\tan\nu_b + K_{bb}) \end{array}\right),
\end{equation}
where $D = (\tan\nu_b + K_{bb})^2 - K_\gamma^2$.
Using this result, we obtain for the spin relaxation probability
 \begin{equation}\label{S}
|S_{{1}/{2} \to -{1}/{2}}|^2  = \frac{K_\gamma^2}{(\tan\nu_b+K_{bb})^2 + \frac{1}{K_1^4} (D-K_1^4)^2}
\end{equation}
This expression immediately yields the scaling law $\sigma_{1/2\to-1/2}\sim \gamma^2$ \cite{Roman03} and illustrates several important features of the mechanism of collisional spin relaxation in $^2\Sigma$ molecules. First, the inelastic probability passes through a maximum at $K_1 = [(\tan\nu_b + K_{bb})^2 -K_\gamma^2]^{1/4}$ and tends to zero as $K_1^{-4}$ when $K_1\to \infty$.  The maximum possible value of $|S_{{1}/{2} \to -{1}/{2}}|^2$  is given by $K_\gamma^2/[\tan\nu_b + K_{bb}]^2$.  This implies that spin relaxation can be slow even when anisotropic interactions are extremely strong ($K_1\to \infty$).
The physical meaning of this result is as follows. In a strongly anisotropic collision system, the interaction potential mixes the incident $N=0$ collision channel with many closed $N>0$ channels. However, spin relaxation cannot occur without the spin-rotation interaction flipping the electron spin within each $N>0$ manifold. Because the spin-rotation interaction is weak, this process is inefficient, acting as a ``dynamical bottleneck'' which is qualitatively similar to the role of the transition state in abstraction chemical reactions \cite{Skodje}.

 In the opposite limit of weakly anisotropic interaction ($K_1\to 0$), Eq. (\ref{S}) yields $|S_{{1}/{2} \to -{1}/{2}}|^2 \sim K_1^4$. This result can also be obtained using third-order distorted-wave approximation  \cite{TBP}. Away from Feshbach resonances, the lowest non-vanishing contribution to the $S$-matrix may be written as $S_{1/2\to -1/2} \sim V_{01}^2 {\gamma}/{B_e^2}$, where $V_{01}$ is the matrix element of the interaction potential between the incident ($N=0$) and closed-channel ($N=1$) distorted waves. The cross sections for spin relaxation thus scale as $V_{01}^4\gamma^2/B_e^4$, the result obtained numerically in \cite{Roman03}.





In summary, we have presented a theoretical analysis of low-temperature collisions of CaH($^2\Sigma$) molecules with Li and Mg atoms based on accurate {\it ab initio} interaction potentials and a rigorous quantum mechanical approach \cite{jcp10}. We have found that the interactions between Li and CaH are extremely strong and anisotropic, but the chemical exchange Li + CaH $\to$ LiH + Ca is not allowed when both collision partners are fully spin-polarized. Our calculations show that inelastic spin relaxation in collisions of fully spin-polarized CaH molecules with Li and Mg atoms occurs slowly, with elastic-to-inelastic ratios in excess of $10^3$ over a wide range of temperatures from 10 $\mu$K to 10 mK. We developed an analytic model, which predicts that spin relaxation cross sections scale as $K_1^4$ with the interaction anisotropy for small $K_1$, but decrease to zero as $K_1\to \infty$. This non-linear scaling should be applicable to all collision-induced relaxation processes involving $^2\Sigma$ molecules at temperatures below the first rotational excitation threshold  ($kT < 2B_e$).



Our results may open up novel avenues of research with cold and ultracold molecules. First, they demonstrate that ultracold spin-polarized mixtures of $^2\Sigma$ molecules with alkali-metal atoms are stable against collisional spin relaxation, strongly suggesting that sympathetic cooling of polar CaH($^2\Sigma$) molecules with Li and Mg atoms will be successful. 
It would be interesting to explore the prospects of sympathetic cooling of laser-cooled SrF($^2\Sigma$) \cite{LaserCooling} and electric-field-decelerated YbF($^2\Sigma$) \cite{YbF}.  Second, our analysis indicates that collisions of $^2\Sigma$ molecules with each other are likely to be predominantly elastic, and hence evaporative cooling of $^2\Sigma$ molecules in a magnetic trap may be feasible.  Finally, our work opens up the possibility of sympathetic cooling of polyatomic spin-1/2 radicals, which can be magnetically trapped at milli-Kelvin temperatures via buffer-gas cooling \cite{polyatomic11}.



We thank H-I. Lu, M. Wright and  J. M. Doyle for discussions and A. Dalgarno and J. F. E. Croft for helpful comments on the manuscript. This work was supported by NSF grants to the Harvard-MIT CUA and ITAMP at Harvard University and the Smithsonian Astrophysical Observatory. A.A.B. and J.K. acknowledge support from RFBR (project No. 03-11-00081) and NSF (Grant No. CHE-0848110) to M.H. Alexander.



\begin{thebibliography}{99}
\bibitem{Constants}
The spectroscopic constants of CaH($^2\Sigma$) are as follows: $B_e=4.205$ cm$^{-1}$, $\gamma=0.0415$ cm$^{-1}$.
\end{thebibliography}
\end{document}